\documentclass[10pt, paper=a4, UKenglish]{article}
\usepackage{graphicx}
\def\Title#1{\begin{center} {\Large #1 } \end{center}}
\def\Author#1{\begin{center}{ \sc #1} \end{center}}
\def\Address#1{\begin{center}{ \it #1} \end{center}}
\def\andauth{\begin{center}{and} \end{center}}

\newcommand\pubblock{\rightline{\begin{tabular}{l} Proceedings of CTD 2020\\ \pubnumber\\
         \pubdate  \end{tabular}}}

\newenvironment{Abstract}{\begin{quotation} \begin{center} 
             \large ABSTRACT \end{center}\bigskip 
      \begin{large}}{\end{large} \end{quotation}}

\newenvironment{Presented}{\begin{quotation} \begin{center} 
             PRESENTED AT\end{center}\bigskip 
      \begin{center}\begin{large}}{\end{large}\end{center} \end{quotation}}

\def\beq{\begin{equation}}
\def\eeq#1{\label{#1}\end{equation}}
\def\eeqn{\end{equation}}

\def\beqa{\begin{eqnarray}}
\def\eeqa#1{\label{#1}\end{eqnarray}}
\def\eeqan{\end{eqnarray}}

\let\bar=\overbar

\def\Dslash{\not{\hbox{\kern-4pt $D$}}}
\def\dslash{\not{\hbox{\kern-2pt $\del$}}}

\def\msb{{\bar{\ssstyle M \kern -1pt S}}}

\usepackage{amsmath, amssymb, amsfonts, authblk, a4wide, bm, cancel, enumerate, appendix, youngtab, verbatim, slashed, subcaption, float, bm}
\usepackage[utf8]{inputenc}
\usepackage[T1]{fontenc}    % use 8-bit T1 fonts
\usepackage{mathtools}
\usepackage[dvipsnames]{xcolor}
\usepackage{forloop}

\usepackage{ifpdf}
\graphicspath{{./img/}}
\usepackage{cleveref}
\usepackage{url}            % simple URL typesetting
\usepackage{booktabs}       % professional-quality tables
\usepackage{nicefrac}       % compact symbols for 1/2, etc.
\usepackage{microtype}      % microtypography
\usepackage{subcaption}     % subfigures

\usepackage{tikz}

\usepackage[style=numeric,sorting=none]{biblatex}

\newcommand\pubnumber{PROC-CTD2020-46}

\newcommand\pubdate{\today}

\def\affiliationA{
Lawrence Berkeley National Laboratory\\
Berkeley, CA, USA}

\def\affiliationB{
Fermi National Accelerator Laboratory \\
Batavia, IL USA}

\def\affiliationC{
California Institute of Technology \\
Pasadena, CA USA}

\def\affiliationD{
University of Cincinnati \\
University of Cincinnati, OH USA}

\def\affiliationE{
Oak Ridge National Laboratory \\
Oak Ridge, TN USA}

\def\affiliationF{
SLAC National Accelerator Laboratory \\
Menlo Park, CA USA}

\newcommand{\conference}{Connecting the Dots Workshop (CTD 2020)\\
April 20-30, 2020}
\usepackage{main-defs}
\usepackage{fancyhdr}
\definecolor{mygrey}{RGB}{105,105,105}
\usepackage{lineno}

\begin{document}

% uncomment the following line for adding line numbers
% \linenumbers

% large size for the first page
\large
\begin{titlepage}
\pubblock

%% Change the title, name, abstract
%% Title 
\vfill
\Title{Track Seeding and Labelling with Embedded-space Graph Neural Networks}
\vfill

%  if you need to add the support use this, fill the \support definition above. 
{\small 

\Author{Nicholas Choma, Daniel Murnane, Xiangyang Ju, Paolo Calafiura, Sean Conlon, \\ Steven Farrell, Prabhat}
\Address{\affiliationA}
\Author{Giuseppe Cerati, Lindsey Gray, Thomas Klijnsma, Jim Kowalkowski, Panagiotis Spentzouris}
\Address{\affiliationB}
\Author{Jean-Roch Vlimant, Maria Spiropulu}
\Address{\affiliationC}
\Author{Adam Aurisano, Jeremy Hewes}
\Address{\affiliationD}
\Author{Aristeidis Tsaris}
\Address{\affiliationE}
\andauth
\Author{Kazuhiro Terao, Tracy Usher}
\Address{\affiliationF}
}
\vfill

\begin{Abstract}
To address the unprecedented scale of HL-LHC data, the Exa.TrkX project is investigating a variety of machine learning approaches to particle track reconstruction. The most promising of these solutions, graph neural networks (GNN), process the event as a graph that connects track measurements (detector hits corresponding to nodes) with candidate line segments between the hits (corresponding to edges). Detector information can be associated with nodes and edges, enabling a GNN to propagate the embedded parameters around the graph and predict node-, edge- and graph-level observables.

Previously, message-passing GNNs have shown success in predicting doublet likelihood, and we here report updates on the state-of-the-art architectures for this task. In addition, the Exa.TrkX project has investigated innovations in both graph construction, and embedded representations, in an effort to achieve fully learned end-to-end track finding.

Hence, we present a suite of extensions to the original model, with encouraging results for hitgraph classification. In addition, we explore increased performance by constructing graphs from learned representations which contain non-linear metric structure, allowing for efficient clustering and neighborhood queries of data points. We demonstrate how this framework fits in with both traditional clustering pipelines, and GNN approaches. The embedded graphs feed into high-accuracy doublet and triplet classifiers, or can be used as an end-to-end track classifier by clustering in an embedded space. A set of post-processing methods improve performance with knowledge of the detector physics. Finally, we present numerical results on the TrackML particle tracking challenge dataset, where our framework shows favorable results in both seeding and track finding.
\end{Abstract}

\vfill

% DO NOT CHANGE!!!
\begin{Presented}
\conference
\end{Presented}
\vfill
\end{titlepage}
\def\thefootnote{\fnsymbol{footnote}}
\setcounter{footnote}{0}
%

% normal size for the rest
\normalsize 

\section{Introduction}

% The physical problem
High energy physics (HEP) experiments are designed to solve some of the most fundamental questions in the universe by probing the interactions of elementary particles in vast quantities of particle collision data. As the frontiers of known physics advance, experiments must increasingly search in regimes of higher energy, higher data volume, and higher data density. In experiments such as ATLAS~\cite{atlas} and CMS~\cite{cms} at the High Luminosity Large Hadron Collider (HL-LHC)~\cite{hllhc}, giant particle detectors will collect measurements from 
% $\mathcal{O}(10^4)$ particles per collision event.
200 particle interactions per collision event on average.
One critical component of the data analysis pipeline in HEP is the reconstruction of charged particle trajectories in high granularity tracking detectors (``particle tracking'').
Tracking systems at the HL-LHC will have $\mathcal{O}(10^8)$ readout channels to record $\mathcal{O}(10^5)$ position measurements (referred to as ``space-points'' or ``hits'') of $\mathcal{O}(10^4)$ charge particles per event.
Tracking algorithms must be able to identify as many of these trajectories as possible while prioritizing the high-transverse momentum particles coming from the highest energy interactions.

% Previous traditional solutions
Traditional tracking solutions in HEP are broken up into several steps. One of these is \textit{seed finding}, in which small combinations of hits (e.g. three-hit ``triplets'') are identified as likely track candidates through hand-crafted criteria. Another, \textit{track building}, involves extrapolating the candidate seeds using Kalman Filters~\cite{kalman-filter} and searching through likely hit candidates at each detector layer until reaching the end of the detector. The combinatorial nature of these algorithms means their computational cost will increase significantly with the expected increase in collision density in the HL-LHC.

% Previous HEP/Exa TrkX works
Motivated by the high computational cost of existing tracking solutions in HEP, the HEP.TrkX pilot project~\cite{heptrkx} and now the Exa.TrkX~\cite{exatrkx} project have been investigating machine learning solutions. Applications using convolutional and recurrent neural networks have been explored but were deemed insufficient to address the challenges of realistic particle tracking~\cite{heptrkx-ctd2017, heptrkx-ctd2018}. Graph neural network (GNN)~\cite{gnn-review1, gnn-review2} models were then proposed and demonstrated to be effective at identifying tracks in realistic data~\cite{heptrkx-ctd2018,exatrkx-ml4ps2019}.
In these applications, graphs are constructed from the point cloud of hits in each event. Edges are drawn between hits that may come from the same particle track according to some loose heuristic criteria. The GNN model is then trained to classify the graph edges as real or fake, giving a pure and efficient sample of track segments which can be used to construct full track candidates.
% then proposed and shown to have real potential as a viable solution

% Contributions and outline of this paper
This work builds off of the previous studies of GNNs applied to particle tracking, advancing in the areas of graph construction and formulation, model performance, and full track reconstruction. All methods are demonstrated on the TrackML dataset~\cite{trkML} using the same preprocessing procedure defined in~\cite{exatrkx-ml4ps2019}, i.e. restricting to the barrel detector only and pre-filtering out the noise hits.
% learned embedding-based graph construction, alternate graph formulations for seeding and improved track finding, and reconstruction of full tracks.
Section~\ref{sec:graph-construction} describes our new approaches for building graphs with learned hit embeddings. Section~\ref{sec:edge-classification} represents the GNN edge classifier and its performance in correctly identifying edges in doublet graphs and triplet graphs, as well as a seeding algorithm transformed from the results of applying the GNN edge classifier on the triplet graphs. Section~\ref{sec:track-labling} shows the track labeling performance of our GNN model. Finally, the conclusion and future work is given in Section~\ref{sec:conclusion}. 
% Section~\ref{sec:edge-classification} bloops and bleeps. Section~\ref{sec:blip} covers our blips. Finally, Section~\ref{sec:pork} is all about the bacon.

% \begin{enumerate}
%     \item The physical problem \cite{acts, atlas, cms, lhc, trkML}
%     \item Previous solutions (including from Hep/Exa TrkX) \cite{DGCNN, interaction-networks}
%     \item Dataset
%     We evaluate the performance of the embedding and GNN on the TrackML dataset \textbf{- this should go earlier, maybe in graph construction?}
%     \item State of the art
%     \item Graphs as natural structure for tracks - GNN description
%     \item How to build graphs - Embedding description
%     \item Paper outline
% \end{enumerate}

\section{Graph Construction}
\label{sec:graph-construction}
We present a general graph construction approach where the objective is to place as many edges as possible between entities that belong together, and as few edges as possible between entities that do not.
In doing so, we first find a good distance metric between pairs of 3D hit measurements, wherein pairs belonging to the same particle are nearby, and pairs belonging to different particles are further apart. 
Assuming the cost to compute the distance between a pair of points is $\mathcal{O}(1)$, we can then construct a sparse graph efficiently by performing neighbor or neighborhood queries.

\subsection{Embedding Architecture}

Rather than learn a distance metric directly, we instead embed our hit measurements $x_i \in \mathcal{X}$ into a new Euclidean space $\mathbb{R}^d$, where $d$ is low enough that the embedded space is not too sparse.
This formulation is an effort to leverage existing frameworks \cite{scikit-learn} which can perform efficient queries using common distance metrics, something we will need for graph construction.

We embed points using a learned model $\psi$, parameterized by $\theta$, which maps points into the new Euclidean space
\begin{equation}
	x \in \mathcal{X} \rightarrow \psi(\theta, x) \in \mathbb{R}^d.
\end{equation}
In our experiments, $x$ includes the 3D hit position in cylindrical coordinates and the shape of the energy deposited by charge particles, $\psi$ is implemented as a multi-layer perceptron (MLP) and $\theta$ are the trainable parameters in MLP.

This stage is trained using a hinge embedding loss, pulling together points belonging to the same particle and pushing apart points which do not.
So for a given sample $s_{ij}=[(x_i,x_j,y_{ij}]$,
where
\begin{equation}
    y_{ij} = 
    \begin{cases}
        1 & \text{if } (x_i,x_j) \text{ belong to the same track}\\
        0 & \text{else},
    \end{cases}
\end{equation}
we compute the loss $l$ as
\begin{equation}
	l(s_{ij}) = \max\left(0, 1-y_{ij} * \left\lVert \psi(\theta, x_i) - \psi(\theta, x_j) \right\rVert_{p}\right).
\end{equation}
For our implementation, we use $p=2$.

Upon completion of the learned embedding's training, we now have a distance metric $d$ where for points $x_i, x_j \in \mathcal{X}$,
\begin{equation}
	d(x_i,x_j) = \left\| \psi(x_i) - \psi(x_j) \right\|_p.
\end{equation}

With $d$, we can construct the input graph $G_{in}$ by querying, for each point $x_i$, the set of points $N_i$ which are nearby.
Then, for every point $x_j \in N_i$, we add a directed edge $e_{i,j}$ which connects node $i$ to node $j$.

For efficient querying, we construct a kd-tree from the embedded points, a binary tree data structure which is constructed in $\mathcal{O}(n \log n)$ time and can be queried in $\mathcal{O}(\log n)$ time.
Once built, each point is queried using one of the following two strategies:
\begin{itemize}
	\item \textbf{$k$-nearest neighbors}, which finds for each point $x_i$ a neighborhood
	\begin{equation}
		N_{knn}^i = knnQuery(x_i, k), |N_i|=k
	\end{equation}
	\item \textbf{$\epsilon$-ball query}, which finds for each point $x_i$ a neighborhood
	\begin{equation}
		N_{ball}^i = ballQuery(x_k, \epsilon),
	\end{equation}
	where for each neighbor 
	\begin{equation}
		x_j \in N_{ball}^i, d(x_i,x_j) < \epsilon.
	\end{equation}
\end{itemize}

Graphs produced with $k$-nearest neighbor queries are regular, thus they allow for grid-like data structures wherein there is no need for sparse matrix multiplication -- something which allows for speedups on GPU. In practice, $\epsilon$-ball queries typically exhibit superior graph construction performance, likely due to the non-uniform density of points in the embedded space.

\begin{figure}[htb]
    \centering
    \includegraphics[width=1.0\textwidth]{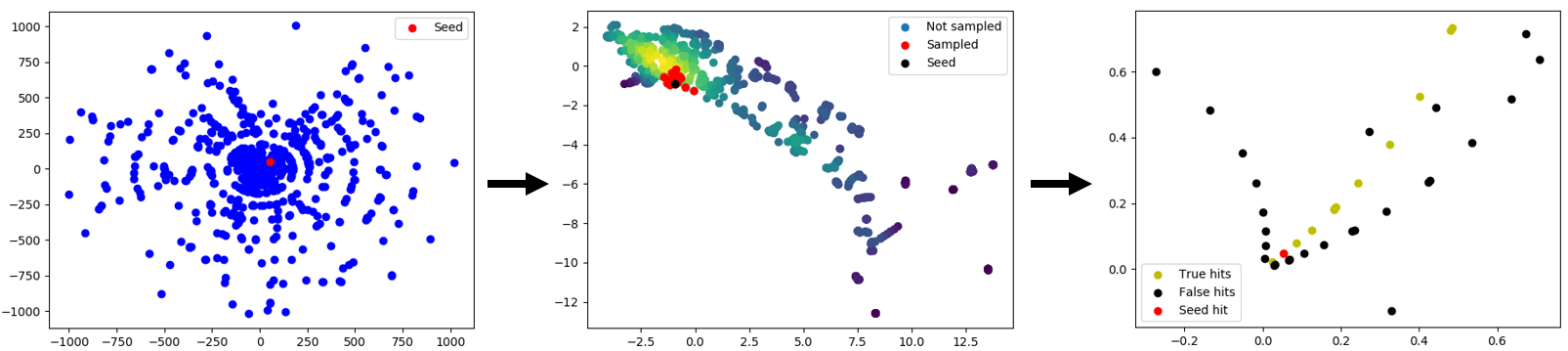}
    \caption{Subset of hits in one event shown in the $xy$ plane with a single hit used to select its corresponding neighborhood (left).
    Selected hits which fall within the seed hit's $\epsilon$-ball radius are shown in a 2d projection of the embedded space (center).
    Selected hits are shown projected back into the original space, and the selected hits which belong to the same track as the seed are shown in yellow (right).
    }
    \label{fig:metric_learning_pipeline}
\end{figure}

Figure \ref{fig:metric_learning_pipeline} shows the process by which neighboring hits are selected from a seeded hit's query to the embedded space.

\subsection{Edge Refinement} Although graphs produced using the learned embeddings are sparse, further refinement can yield still much sparser graphs.
Within the embedding model, we are only able to consider features derived from each point individually.
Since we have now produced a relatively small set of edges, represented as pairs of points, we can now consider models which take as input pairs of points, as well as pairwise features derived from domain expertise.

We thus construct an edge refinement model $\psi'$, parameterized by $\theta'$, which operates on pairs of points $x_i, x_j$ and their pairwise features $z_{ij}$, and outputs the probability $p_{ij}$ that the pair belongs to the same cluster.
\begin{equation}
	p_{ij} = \psi'(x_i,x_j,z_{ij}) \in [0,1]
\end{equation}
$\psi'$ is likewise parameterized as a multi-layer preceptron.

With our trained model, we compute $p_{ij}$ for each $e_{ij} \in E$ produced during the embedding stage.
Then, choosing a threshold hyperparameter $t \in [0,1]$, we are left with our final edge selection
\begin{equation}
	E' = \{e_{ij} | p_{ij} > t\}.
\end{equation}

\subsection{Performance}

To achieve competitive performance with traditional tracking algorithms, the graph construction stage must run in approximately one second or less while maintaining a sufficiently high portion of the graph's true edges.
Whereas the embedding model $\psi$ must only consider $N_i$ points, the edge refinement model $\psi'$ must infer over $|E_i|$ pairs of points and as such acts as a bottleneck.
To mediate this bottleneck, $\psi'$ is a relatively small network containing just 3 hidden layers with 512 hidden units each.
Additionally, $\psi'$ uses half-precision parameters which is able to achieve a 2x speedup over full precision when run on Nvidia's GPU architectures.

We also note the adaptability of our architecture to differing edge recovery and graph size requirements through the neighborhood and filtering hyperparameters, $\epsilon$ and $t$, respectively.
In our tests, we required 96\% of the true edges to be recovered by the graph construction pipeline to maintain a high TrackML score. 
Respecting the timing requirements for this stage, our architecture was thus capable of graph construction where 30.3\% of all edges were true edges.
This result has significant implications for downstream GNN training and inference, allowing for vastly reduced computation in graph convolution, and a smaller memory footprint during training which eliminates the need to divide the domain onto multiple GPUs.

%\textbf{Dataset}

%We evaluate the performance of the embedding and GNN on the TrackML dataset \textbf{- this should go earlier, maybe in graph construction?}

\section{GNN Edge Classification}
\label{sec:edge-classification}

\subsection{GNN Architecture}

We extend the prototypical message passing Graph Neural Network architecture as described in \cite{kipf-welling}, with an attention mechanism \cite{velivckovic2017graph} and a ResNet-style \cite{he2016deep} addition operation between message passing layers to help reduce the vanishing gradient effect. Once hits are assembled into input graphs $G_{in}$ in embedded space (\cref{sec:graph-construction}), the hit coordinates of the $i^{th}$ node $x_i = (r_i, \phi_i, z_i)$ are passed through a $(n_{\text{layers,in}}, n_{\text{dims,hidden}})$ input MLP, where $n_\text{layers,in}$ are the number of fully connected layers between cylindrical coordinates and the latent node features $h_i$, and $n_\text{dims,in}$ is the width of these layers (generally, we take MLPs as having the same number of parameters in each layer). 

We then include a recurrent set of $\{EdgeNetwork, NodeNetwork\}$, iterating $n_\text{iter}$ times through~(\cref{fig:architecture}). In its forward pass, the $EdgeNetwork$ concatenates the $n_\text{dims,hidden}$ features of the nodes on either end of each edge and passes this through a $(n_\text{layers,edge}, n_\text{dims,hidden})$ edge MLP, with one fully connected layer outputting a scalar value for each edge. For an edge connecting nodes $i$ and $j$, this value is called the \textit{edge score} $s_{ij}$, defined for the $l^{th}$ iteration as
\begin{align}
    s_{ij}^{l} = \text{MLP}( h_i^l \oplus h_j^l)
\end{align}
where $\oplus$ is a concatenation of the hidden features, and MLP is a sequence of multiplications by weight matrices and operations of non-linear activations, in this case Tanh functions.
This edge score is used in an attention mechanism. The $NodeNetwork$ implements a message passing forward pass, such that for each node, the neighboring node features of all incoming edges are aggregated with a weighted sum. The same is done with outgoing edges, then these two pooled vectors are concatenated and passed through a $(n_\text{layers,node}, n_\text{dims,hidden})$ node MLP. 
\begin{align}
    h_i^{l+1} = \text{MLP}( \sum\limits_j s^{in}_{ij} \cdot h_j^l \; \oplus \;  \sum\limits_j s^{out}_{ij} \cdot h_j^l \; \oplus \; h_i^l) + h_i^l
\end{align}
As can be seen, the the output of the MLP is summed with the hidden features of the previous iteration (a "skip connection"). After $n_\text{iter}$ iterations, the node features are passed through the Edge Network a final time, to determine the edge scores, which are interpreted as a truth likelihood and handled by a binary cross entropy loss function.

\hspace{-1cm}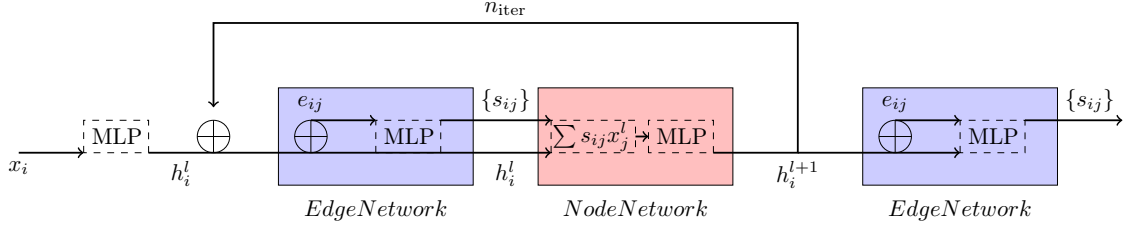
\begin{figure}
    \centering
    \resizebox{\linewidth}{!}{%
    \begin{tikzpicture}
        % \draw[step=1cm,gray,very thin] (0,0) grid (15,6);
       
       % Network boxes
        \draw [dashed] (-1,1) rectangle (0, 1.5);
        \node [right] at (-1,1.25) {MLP};
        \draw [fill=blue, fill opacity=0.2] (2,0.5) rectangle (5,2);
        \draw (2.5,1.25) circle (.25);
        \draw (2.5,1) -- (2.5,1.5);
        \draw (2.25,1.25) -- (2.75,1.25);
        \node [above] at (2.5, 1.5) {$e_{ij}$};
        \draw [dashed] (3.5,1) rectangle (4.5,1.5);
        \node [right] at (3.5, 1.25) {MLP};
        \draw [fill=pink, fill opacity=1] (6,0.5) rectangle (9,2);
        \node [right] at (6.1,1.25) { $\sum s_{ij} x_j^l$};
        \draw [dashed] (6.2,1) rectangle (7.5,1.5);
        \draw [dashed] (7.7,1) rectangle (8.7,1.5);
        \node [right] at (7.7, 1.25) {MLP};
        \draw[thick, ->] (7.5,1.25) -- (7.7,1.25);
        
        \draw [fill=blue, fill opacity=0.2] (11,0.5) rectangle (14,2);
        \draw (11.5,1.25) circle (.25);
        \draw (11.5,1) -- (11.5,1.5);
        \draw (11.25,1.25) -- (11.75,1.25);
        \node [above] at (11.5, 1.5) {$e_{ij}$};
        \draw [dashed] (12.5,1) rectangle (13.5,1.5);
        \node [right] at (12.5, 1.25) {MLP};
        
        % Feature flow
        \draw[thick, ->] (-2,1) -- (-1,1);
        \draw[thick, ->] (0,1) -- (6.2,1);
        \draw[thick, ->] (8.7,1) -- (10,1) -- (10,3) -- (1,3) -- (1,1.7);
        \draw[thick, ->] (10,1) -- (12.5,1);
        \draw[thick, ->] (4.5, 1.5) -- (6.2, 1.5);
        \draw[thick, ->] (2.5, 1.5) -- (3.5, 1.5);
        \draw[thick, ->] (11.5, 1.5) -- (12.5, 1.5);
        \draw[thick, ->] (13.5, 1.5) -- (15, 1.5);
        \node [below] at (-2,1) {$x_i$};
        \node [below] at (0.5,1) {$h_i^l$};
        \node [below] at (5.5,1) {$h_i^l$};
        \node [below] at (10,1) {$h_i^{l+1}$};
        \node [above] at (5.5,1.5) {$\{s_{ij}\}$};
        \node [above] at (14.5,1.5) {$\{s_{ij}\}$};
        \node [below] at (3.5, 0.4) {$EdgeNetwork$};
        \node [below] at (7.5, 0.4) {$NodeNetwork$};
        \node [below] at (12.5, 0.4) {$EdgeNetwork$};
        \node [above] at (5.5, 3) {$n_\text{iter}$};

        % Concatenate
        \draw (1,1.25) circle (.25);
        \draw (1,1) -- (1,1.5);
        \draw (0.75,1.25) -- (1.25,1.25);
    \end{tikzpicture}}
    \caption{Recurrent Attention Message Passing Architecture}
    \label{fig:architecture}
\end{figure}

In order to determine the best set of model hyperparameters, we perform Bayesian optimisation over them, with the goal of optimising both edge classification efficiency and purity. In practice, we aim to maximise the value $f_1 = 2 \cdot \frac{\text{eff} \times \text{pur}}{\text{eff} + \text{pur}} \in (0,1)$ introduced in \cite{10.3115/1072064.1072067}, where
\begin{align}
    \text{eff} = \frac{n_\text{true,positive}}{n_\text{true}}, && \text{pur} = \frac{n_\text{true,positive}}{n_\text{positive}}.
\end{align}

% Details of the hyperparameter scan are given in \cref{tab:hpo}, for both doublet and triplet GNNs.

% \begin{table}[]
%     \centering
%     \begin{tabular}{@{}llllllll@{}} \toprule 
%          & \multicolumn{3}{c}{Doublet GNN} & \phantom{abc} & \multicolumn{3}{c}{Triplet GNN} \\
%         \cmidrule{2-4} \cmidrule{6-8}
%         H.P.s & Min & Max & Best && Min & Max & Best \\ \midrule 
%         $n_\text{iter}$ & 1 & 9 & 6 && 1 & 9 & 4 \\
%         $X$ & 1 & 2 & 3 && 1 & 2 & 3 \\
%         $X$ & 1 & 2 & 3 && 1 & 2 & 3 \\
%         $X$ & 1 & 2 & 3 && 1 & 2 & 3 \\
%         $X$ & 1 & 2 & 3 && 1 & 2 & 3 \\
%         $X$ & 1 & 2 & 3 && 1 & 2 & 3 \\
%         \bottomrule
%     \end{tabular}
%     \caption{Summary of hyperparameter (H.P.) Bayesian optimisation, \red{could go in an appendix}}
%     \label{tab:hpo}
% \end{table}

\subsection{Doublet GNN Performance}

\begin{figure}[htb!]
    \centering
    \begin{subfigure}[b]{0.9\linewidth}
        \centering
        \includegraphics[width=1\textwidth]{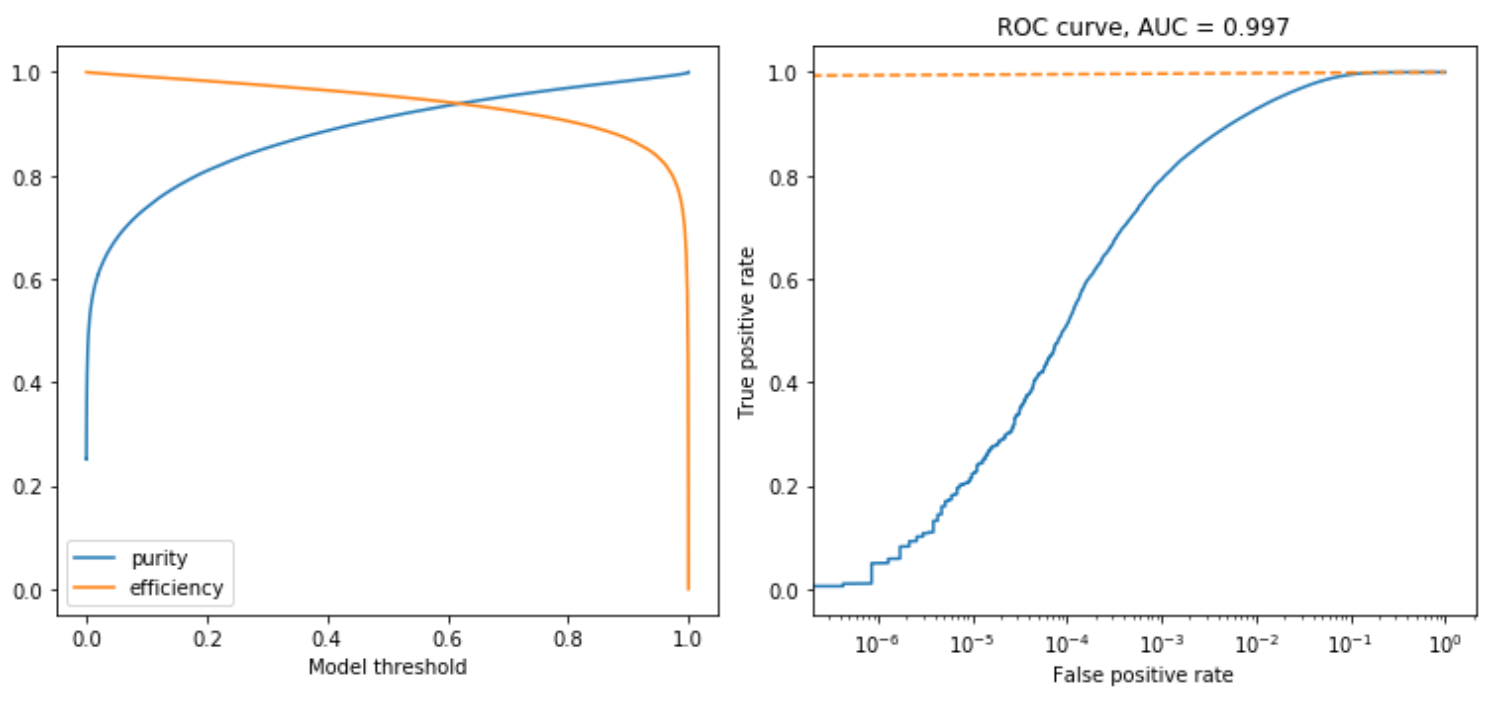}\caption{Doublet GNN}
    \end{subfigure} 
    \begin{subfigure}[b]{0.9\linewidth}
        \centering
        \includegraphics[width=1\textwidth]{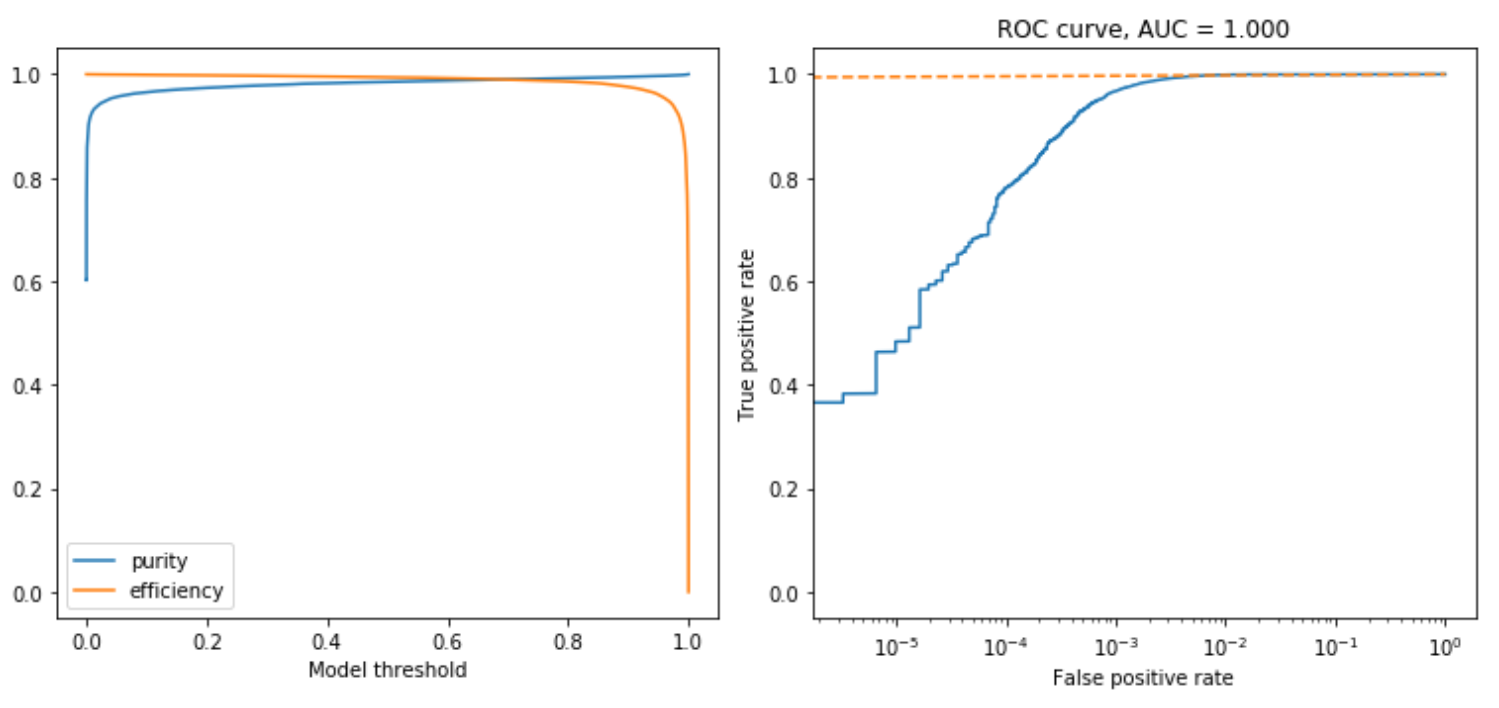}\caption{Triplet GNN}
    \end{subfigure} 
    \caption{GNN performance metrics}
    \label{fig:GNN_results}
\end{figure}

Given the above architecture, we present the results of edge classification. \Cref{fig:GNN_results} gives the efficiency and purity at different choices of edge score threshold (ROC curve). The ROC area under curve (AUC) for the best doublet GNN hyperparameter configuration is 0.997. As a matter of memory management, the hit graphs must be split into subgraphs. We find that 8 subgraphs, segmented around the $\phi$-direction. To preserve edges, each full graph is first constructed, hits in each $\phi$ slice are assigned to subgraphs, as are copies of hits connected by an edge to those hits. 

\subsection{Triplet GNN Performance}\label{sec:triplet_gnn}

To perform classification of triplets using the same approach as the doublet classification, we need to identify hitgraph doublets as nodes in a new "triplet graph", and combinations of triplets as edges in a triplet graph. To accelerate this transformation, we first convert the edge list (the standard Pytorch Geometric COO format \cite{2019arXiv190302428F}) to sparse row matrices (CSR format) on a GPU with CuPy \cite{Okuta2017CuPyA}. These two matrices, one incoming, on outgoing, are multiplied to produce a square CSR matrix that represents triplet edges. That is,
\begin{align}
    (e^{triplet}_{CSR})_{ij} = (e^T_{CSR,out} \times e_{CSR,in})_{ij} = 
    \begin{cases}
        0, & \text{if  } e_i^{out} \neq e_j^{in}\\
        1, & \text{if  } e_i^{out} = e_j^{in}
    \end{cases}, \;\; \text{where} \;\;
    (e_{CSR})_{ij} := 
    \begin{cases}
        0, & \text{if  } e_i^{out} \neq i\\
        1, & \text{if  } e_i^{out} = i
    \end{cases}
\end{align}

% \begin{align}
%     & e_{COO} = \begin{pmatrix}
%     e_0^{in} & e_1^{in} & ... & e_n^{in}\\
%     e_0^{out} & e_1^{out} & ... & e_n^{out}
%     \end{pmatrix}
% \end{align}

% \begin{align}
%     \rightarrow \hspace{2em} & e_{CSR,in} = 
%     {\tiny \begin{matrix}
%         & \begin{matrix} 0 & 1 & \cdots & n \end{matrix}\\
%         \begin{matrix}
%         \cdots \\
%         e_0^{in}\\
%         \cdots \\
%         e_1^{in}\\
%         \cdots \\
%         e_n^{in}\\
%         \cdots
%         \end{matrix} & \begin{pmatrix}
%         \cdots \\
%         1 & 0 & \cdots & 0 \\
%         \cdots \\
%         0 & 1 & \cdots & 0 \\
%         \cdots \\
%         0 & 0 & \cdots & 1 \\
%         \cdots
%         \end{pmatrix}
%     \end{matrix} }, && \text{and} && e_{CSR,out}^T {\tiny = 
%     \begin{matrix}
%         &  \setlength\arraycolsep{2pt}  \begin{matrix} ... & e_0^{out} & ... & e_1^{out} & ... & e_n^{out} & ... \end{matrix}\\
%         \begin{matrix}
%         0 \\
%         1 \\
%         \vdots \\
%         n
%         \end{matrix} & \begin{pmatrix}
%         ... & 1 & ... & 0 & ... & 0 & ... \\
%         ... & 0 & ... & 1 & ... & 0 & ... \\
%         \vdots \\
%         ... & 0 & ... & 0 & ... & 1 & ...
%         \end{pmatrix}
%     \end{matrix} }
% \end{align}

This efficient transformation is able to decrease the time taken for each event, from the inbuilt methods of Pytorch Geometric of $\mathcal{O}(1s)$, to $\mathcal{O}(100ms)$, thereby making the prospect of sub-second triplet classification possible. Once the triplet graph is constructed, the same GNN architecture of the previous section is used in training and edge classification. Node features in the triplet graph are defined by concatenating the node features of each doublet, along with the classification score of the associated edge, such that for two nodes $i$ and $j$ connected by an edge $e_{ij}$ 
\begin{align}
    x^{triplet}_e = (r_i, \phi_i, z_i) \oplus (r_j, \phi_j, z_j) \oplus s_{ij}.
\end{align}
A cut is placed on the edges used in the triplet graph construction, so as to limit combinatorial growth. Cutting doublet edges below a score of $s_{ij} < 0.1$ boosts the graph purity from ~30\% to ~60\%, while retaining 99.12\% efficiency. Training the triplet GNN on the hyperparameter configuration given above produces the performance given in \cref{tab:GNN_performance}.

\begin{table}[htb]
    \centering
    \begin{tabular}{@{}llllll@{}} \toprule 
         & \multicolumn{2}{c}{Doublet GNN} & \phantom{abc} & \multicolumn{2}{c}{Triplet GNN} \\
        \cmidrule{2-3} \cmidrule{5-6}
        Threshold & 0.5 & 0.8 && 0.5 & 0.8 \\ \midrule 
        Accuracy & 0.9761 & 0.9784 && 0.9960 & 0.9957\\
        Purity & 0.9133 & 0.9694 && 0.9854 & 0.9923 \\
        Efficiency & 0.9542 & 0.9052 && 0.9939 & 0.9850 \\
        \bottomrule
    \end{tabular}
    \caption{Performance of doublet and triplet GNNs at given thresholds.}
    \label{tab:GNN_performance}
\end{table}

\begin{figure}[htb]
    \centering
    \includegraphics[width=0.5\linewidth]{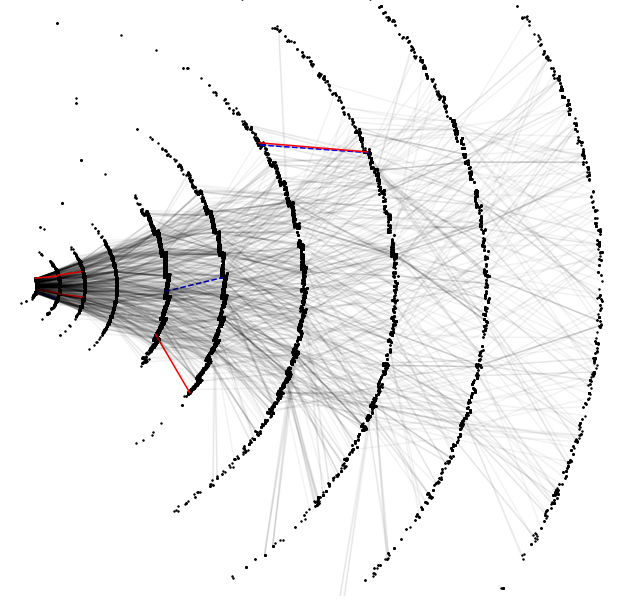}
    \caption{Edge classification on example hitgraph. Black: True positive (transparent for clarity), blue: False negative, red: false positive.}
    \label{fig:seed_performance}
\end{figure}

\begin{figure}[htb]
    \centering
    \includegraphics[width=0.48\textwidth]{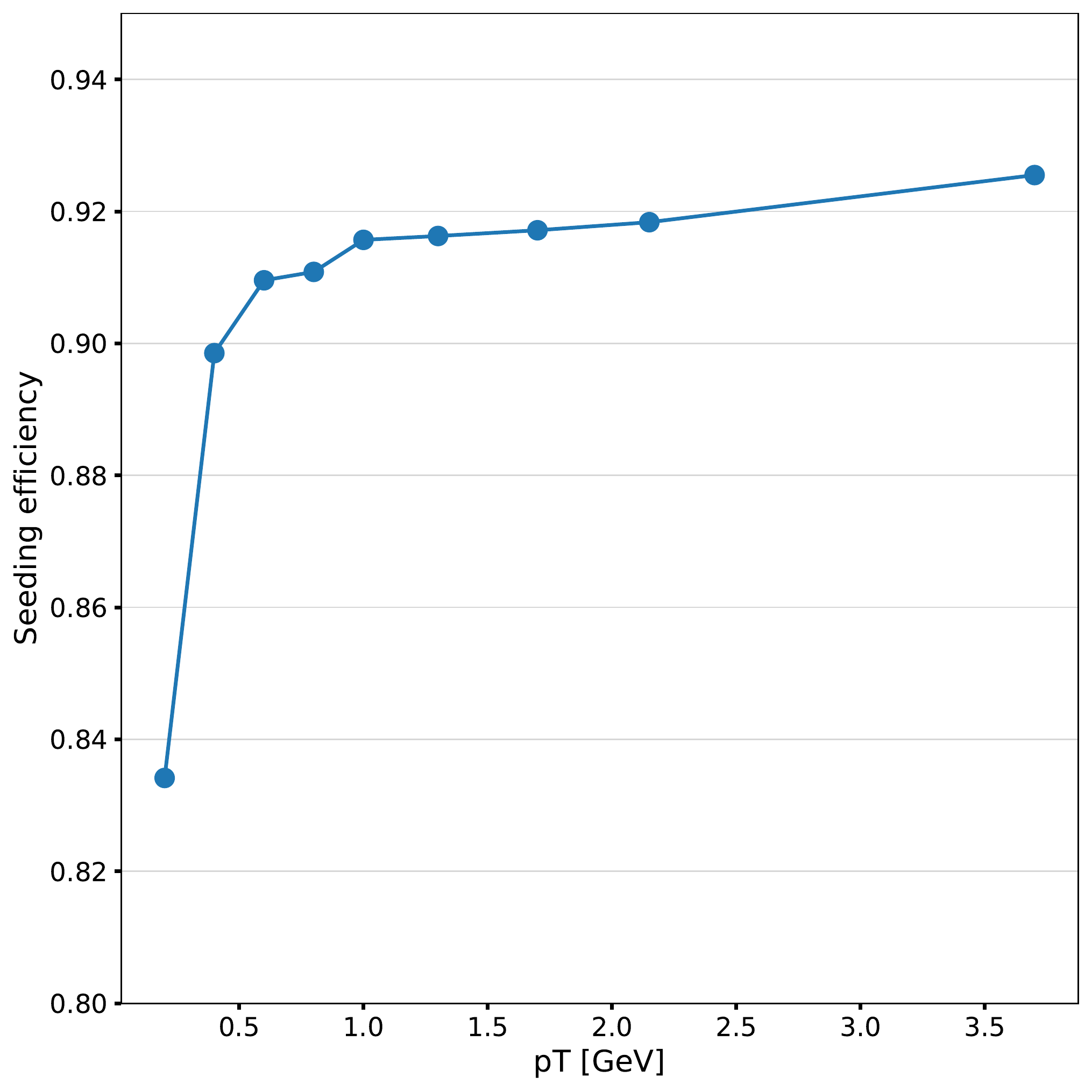}
    \includegraphics[width=0.48\textwidth]{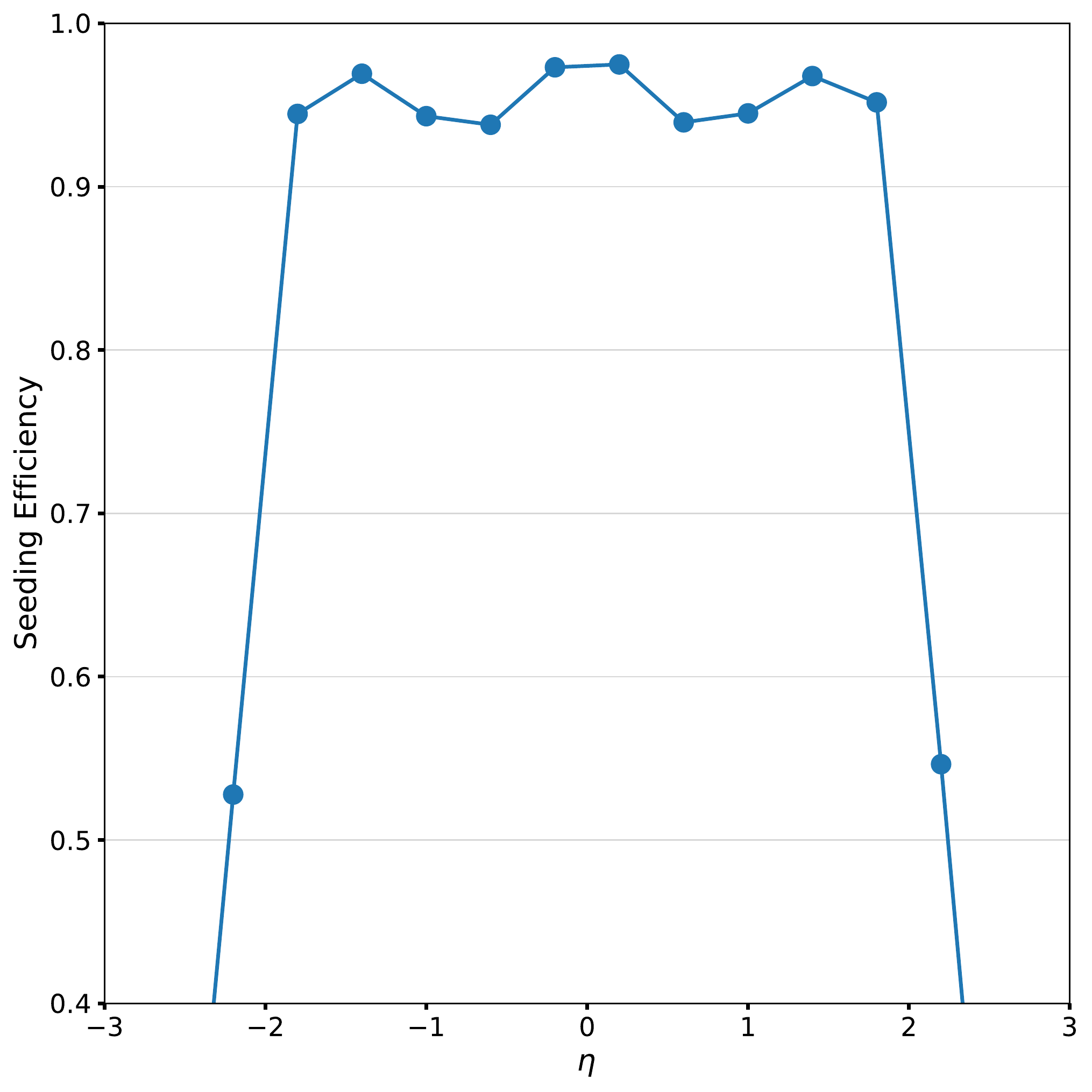}
    \caption{Seeding efficiency of the triplet-GNN-based seeding algorithm, defined as the ratio of the number of good tracks matched to at least one seed candidate over the total number of good tracks, as a function of  $p_{\text{T}}$ (left) and $\eta$ (right). 
    Good tracks are defined as the tracks that are resulted from particles leaving at least three hits in different layers and having at least five hits in the triplet graph.
    }
    \label{fig:seed_performance2}
\end{figure}

Seeds, defined as a set of at least three hits recorded by consecutively different layers,
are the crucial inputs for the existing tracking reconstruction algorithms~\cite{Cornelissen_2008}.
The triplet GNN was turned into a seeding algorithm in which the edges with a high GNN score are selected and the nodes connecting each edge form a seed candidate. 
The performance of the GNN-based seeding algorithm is evaluated in terms of seeding efficiency, defined as the ratio of the number of good tracks matched to at least one seed candidate over the total number of good tracks, 
and seeding purity, defined as the number of seed candidates matched to a good track over the total number of seed candidates.
Good tracks are defined as the tracks that are resulted from particles leaving at least three hits in different layers and having at least five hits in the triplet graph. 
Evaluated on 100 testing events, the GNN-based seeding algorithm renders 
a seeding efficiency of ($88.6 \pm 0.2$)\% and a seeding purity of ($99.08 \pm 0.07$)\%.
Only statistical uncertainties are taken into account.
The seeding efficiency is further evaluated as a function of the transverse momentum ($p_{\text{T}}$) and the pseudo-rapidity~\footnote{$\eta = -\ln\tan(\theta/2)$, where $\theta$ is the polar angle.} ($\eta$) of the particle that the track is associated with, shown in Figure~\ref{fig:seed_performance2}. The GNN-based seeding algorithm has an efficiency of 83\% for particles of  $p_{\text{T}}$\ in [0.1, 0.3]~\gev\ and increases to 92\% for particles with $p_{\text{T}}$\ at or above 0.7~\gev.

% \textbf{Triplet seed comparison with existing results}, \cref{fig:seed_comparison}.

\section{Track Labeling Performance}
\label{sec:track-labling}

\begin{figure}[htb]
    \centering
    \includegraphics[width=1\linewidth]{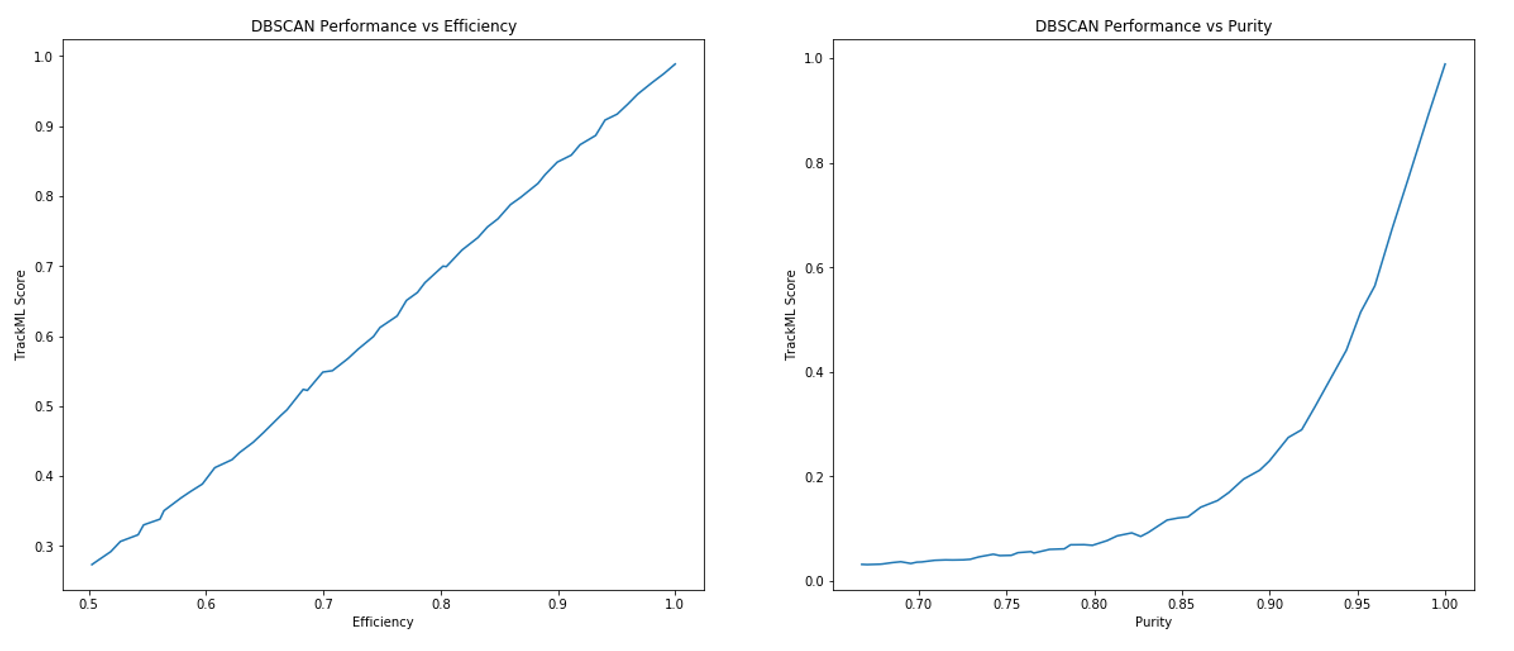}
    \caption{The performance of DBSCAN against generic efficiency/purity scaling}
    \label{fig:DBSCAN_performance}
\end{figure}

Given a graph of classified (doublet or triplet) edges, we would like to use these scores $e_{ij}$ to assign unique track labels to each hit. The approach we use here is to apply DBSCAN (Density-Based Spatial Clustering of Applications with Noise) \cite{scikit-learn}, with $\textit{min\_samples}=1$. Recent releases of DBSCAN allow a sparse matrix as a precomputed metric input. In practice, we take the COO-format edge list, convert it to a CSR-format sparse matrix as described in \cref{sec:triplet_gnn}, and assign each entry a distance $d_{ij}$, defined as $d_{ij} = 1 - e_{ij}$. The \textit{neighborhood distance} $\epsilon$ is left as a hyperparameter to be tuned for the best track labelling performance. This performance we measure according to the \textit{TrackML Score} as defined in the TrackML Challenge Kaggle Competition (\textbf{cite}). The score $ s \in (0, 1)$ is a weighted sum of each correctly labelled hit, giving more importance to straighter and longer tracks, and particularly hits at the beginning and end of these tracks that could be used as seeds. DBSCAN outputs integer cluster labels, which are used directly for calculating the TrackML Score against the truth labels. For a graph created from truth, with efficiency and purity artificially tuned, the TrackML score produced with the DBSCAN method scales as in \cref{fig:DBSCAN_performance}. We see that, provided purity is close to 100\%, DBSCAN will generally deliver a faithful score. The score produced with this method drops exponentially with purity, but is robust (dropping linearly) against inefficiency. There are methods that will cluster more robustly for drops in efficiency and purity, and these should be explored in future works. In this work, we settle on DBSCAN for its simplicity and fast performance, being careful to note when a drop in score is merely an artifact of DBSCAN or some more intrinsic failure of the GNN classification or embedding construction.

\begin{table}[htb]
    \centering
    \begin{tabular}{@{}lll@{}} \toprule 
         & \multicolumn{2}{c}{TrackML Score} \\
        \cmidrule{2-3}
        Condition & Truth & Prediction \\ \midrule 
        Doublet GNN & 0.935 & 0.805\\
        Triplet GNN & 0.846 & 0.815\\
        \; \; and $\eta \in (-2.1, 2.1)$ & 0.912 & 0.876 \\
        \; \; and 5 hits adjacent & - & 0.932 \\
        \bottomrule
    \end{tabular}
    \caption{TrackML score for each stage of pipeline and condition.}
    \label{tab:trackml_scores}
\end{table}

We calculate the TrackML score against various possible conditions on the dataset. Each condition and its corresponding maximum score is given in \cref{tab:trackml_scores}. As we are only classifying hits in the barrel, we normalise against this condition. The table gives the maximum score attainable with the edges provided from the metric learning neighbourhood construction stage, and applying DBSCAN to truth-level classification. This stems from the ~$96\%$ efficiency of the construction stage, leading to a ~$6.5\%$ loss in TrackML score. This is consistent with the generic scaling seen in \cref{fig:DBSCAN_performance}. The actual performance of the doublet GNN is given as $s=0.805$. 

The maximum score attainable with the triplets constructed from the doublet classification stage (as a reminder, all hits connected in a graph are constrained to adjacent layers in the detector), again using truth-level classification is $s=0.846$. This is another large reduction in possible score, this time from heavily-weighted doublets that are not included in the triplet construction as they are not joined by likely edges. These are predominantly at the edges of the barrel, where they are part of a track dominated by endcap layer hits. By narrowing the pseudo-rapidity range of possible hits to $\eta \in (-2.1, 2.1)$ and removing a small number of "fragments" (these tails of longer tracks in the endcaps) we reclaim much of the maximum possible score lost in the triplet construction. Finally, given that we artificially restrict our study to adjacent layers, we restrict tracks with greater than five hits to contain at least five adjacent hits, giving the final adjusted score of $s=0.932$. 

\section{Conclusion}
\label{sec:conclusion}

The pipeline presented here represents a significant improvement in track labelling and seeding performance. To apply the stages of preprocessing, KNN clustering in learned embedded space, pair filtering, GNN classification of doublets and triplets, then either seed generation or track finding requires $\mathcal{O}(5s)$ per event. Restricting the pseudorapidity to focus on the barrel ($\eta \in (-2,2)$), we have seed efficiency $>93\%$ and purity $>99\%$, while the track finding gives a TrackML score of $0.932$ given reconstructability constraints. These metrics compare favourably with traditional methods of seeding and track finding, and moreover allow for fast performance and parallelizability -- features often lacking due to the scaling problems inherent in many traditional algorithms.

We note that several artificial advantages were incorporated into the work, including ignoring noise hits and excluding data from the endcaps. Current work is focused on incorporating that data back into the classification pipeline and further advancing the computational and physics performance of the models, including testing our pipeline on data simulated by HL-LHC experiments. We will also study the robustness of our solution against various systematical effects such as detector noise and misalignment. To meet the requirements of HL-LHC tracking we need to improve the physics and computational performance of our models. To this end, we are exploring the utilization of more advanced GNN architectures, next-generation GP-GPUs and Google Cloud TPUs, and of distributed training and inference.

\subsubsection*{Software Availability}
The software and the documentation needed to reproduce the results of this article are available at \url{https://github.com/exatrkx/exatrkx-ctd2020}

\section*{Acknowledgments}

This research was supported in part by the Office of Science, Office of High Energy Physics, of the US Department of Energy under Contracts No. DE-AC02-05CH11231 (CompHEP Exa.TrkX) and No. DE-AC02-07CH11359 (FNAL LDRD 2019.017).

This research used resources of the National Energy Research Scientific Computing Center (NERSC), a U.S. Department of Energy Office of Science User Facility operated under Contract No. DE-AC02-05CH11231.

\printbibliography

@article{hllhc,
      author         = "Apollinari, G. and Brüning, O. and Nakamoto, T. and
                        Rossi, Lucio",
      title          = "{High Luminosity Large Hadron Collider HL-LHC}",
      journal        = "CERN Yellow Report",
      year           = "2015",
      number         = "5",
      pages          = "1-19",
      doi            = "10.5170/CERN-2015-005.1",
      eprint         = "1705.08830",
      archivePrefix  = "arXiv",
      primaryClass   = "physics.acc-ph",
      reportNumber   = "FERMILAB-DESIGN-2017-02",
      SLACcitation   = "%%CITATION = ARXIV:1705.08830;%%"
}

@article{atlas,
      author         = "Aad, G. and others",
      title          = "{The ATLAS Experiment at the CERN Large Hadron Collider}",
      collaboration  = "ATLAS",
      journal        = "JINST",
      volume         = "3",
      year           = "2008",
      pages          = "S08003",
      doi            = "10.1088/1748-0221/3/08/S08003",
      SLACcitation   = "%%CITATION = JINST,3,S08003;%%"
}

@article{cms,
      author         = "Chatrchyan, S. and others",
      title          = "{The CMS experiment at the CERN LHC}",
      collaboration  = "CMS",
      journal        = "JINST",
      volume         = "3",
      year           = "2008",
      pages          = "S08004",
      doi            = "10.1088/1748-0221/3/08/S08004",
      SLACcitation   = "%%CITATION = JINST,3,S08004;%%"
}

@article{heptrkx-ctd2017,
      author         = "Farrell, Steven and others",
      title          = "{The HEP.TrkX Project: deep neural networks for HL-LHC
                        online and offline tracking}",
      booktitle      = "{Proceedings, Connecting The Dots / Intelligent Tracker
                        (CTD/WIT 2017): Orsay, France, March 6-9, 2017}",
      journal        = "EPJ Web Conf.",
      volume         = "150",
      year           = "2017",
      pages          = "00003",
      doi            = "10.1051/epjconf/201715000003",
      reportNumber   = "FERMILAB-CONF-17-326-CD",
      SLACcitation   = "%%CITATION = 00776,150,00003;%%"
}

@inproceedings{heptrkx-ctd2018,
      author         = "Farrell, Steven and others",
      title          = "{Novel deep learning methods for track reconstruction}",
      booktitle      = "{4th International Workshop Connecting The Dots 2018
                        (CTD2018) Seattle, Washington, USA, March 20-22, 2018}",
      year           = "2018",
      eprint         = "1810.06111",
      archivePrefix  = "arXiv",
      primaryClass   = "hep-ex",
      reportNumber   = "FERMILAB-CONF-18-598-CD",
      SLACcitation   = "%%CITATION = ARXIV:1810.06111;%%"
}

@article{gnn-review1,
  author    = {Jie Zhou and Ganqu Cui and Zhengyan Zhang and
               Cheng Yang and Zhiyuan Liu and
               Lifeng Wang and Changcheng Li and Maosong Sun},
  title     = {Graph Neural Networks: A Review of Methods and Applications},
  journal   = {},
  volume    = {abs/1612.00222},
  year      = {2019},
%   url       = {http://arxiv.org/abs/1812.08434},
  archivePrefix = {arXiv},
  eprint    = {1812.08434},
%   biburl    = {https://dblp.org/rec/bib/journals/corr/BattagliaPLRK16},
%   bibsource = {dblp computer science bibliography, https://dblp.org}
}

@article{gnn-review2,
  author    = {Michael M. Bronstein and
               Joan Bruna and
               Yann LeCun and
               Arthur Szlam and
               Pierre Vandergheynst},
  title     = {Geometric deep learning: going beyond Euclidean data},
  journal   = {CoRR},
  volume    = {abs/1611.08097},
  year      = {2016},
%   url       = {http://arxiv.org/abs/1611.08097},
  archivePrefix = {arXiv},
  eprint    = {1611.08097},
  timestamp = {Mon, 13 Aug 2018 16:48:20 +0200},
  biburl    = {https://dblp.org/rec/bib/journals/corr/BronsteinBLSV16},
  bibsource = {dblp computer science bibliography, https://dblp.org}
}

@misc{trkML,
	title="{TrackML Particle Tracking Challenge}",
	url="https://www.kaggle.com/c/trackml-particle-identification"
}

@article{Cornelissen_2008,
	doi = {10.1088/1742-6596/119/3/032014},
	url = {https://doi.org/10.1088%2F1742-6596%2F119%2F3%2F032014},
	year = 2008,
	month = {7},
	publisher = {{IOP} Publishing},
	volume = {119},
	number = {3},
	pages = {032014},
	author = {T Cornelissen and M Elsing and I Gavrilenko and W Liebig and E Moyse and A Salzburger},
	title = {The new {ATLAS} track reconstruction ({NEWT})},
	journal = {Journal of Physics: Conference Series}
}

@article{scikit-learn,
 title={Scikit-learn: Machine Learning in {P}ython},
 author={Pedregosa, F. and Varoquaux, G. and Gramfort, A. and Michel, V.
         and Thirion, B. and Grisel, O. and Blondel, M. and Prettenhofer, P.
         and Weiss, R. and Dubourg, V. and Vanderplas, J. and Passos, A. and
         Cournapeau, D. and Brucher, M. and Perrot, M. and Duchesnay, E.},
 journal={Journal of Machine Learning Research},
 volume={12},
 pages={2825--2830},
 year={2011}
}

@inproceedings{exatrkx-ml4ps2019,
    author = "Ju, Xiangyang and others",
    title = "{Graph Neural Networks for Particle Reconstruction in High Energy Physics detectors}",
    booktitle = "{33rd Annual Conference on Neural Information Processing Systems}",
    eprint = "2003.11603",
    archivePrefix = "arXiv",
    primaryClass = "physics.ins-det",
    reportNumber = "FERMILAB-CONF-20-163-PPD-QIS-SCD",
    month = "3",
    year = "2020"
}

@article{kipf-welling,
  author    = {Thomas N. Kipf and
               Max Welling},
  title     = {Semi-Supervised Classification with Graph Convolutional Networks},
  journal   = {CoRR},
  volume    = {abs/1609.02907},
  year      = {2016},
  url       = {http://arxiv.org/abs/1609.02907},
  archivePrefix = {arXiv},
  eprint    = {1609.02907},
  bibsource = {dblp computer science bibliography, https://dblp.org}
}

@article{velivckovic2017graph,
  title={Graph attention networks},
  author={Veli{\v{c}}kovi{\'c}, Petar and Cucurull, Guillem and Casanova, Arantxa and Romero, Adriana and Lio, Pietro and Bengio, Yoshua},
  journal={arXiv preprint arXiv:1710.10903},
  year={2017}
}

@ARTICLE{2019arXiv190302428F,
       author = {{Fey}, Matthias and {Lenssen}, Jan Eric},
        title = "{Fast Graph Representation Learning with PyTorch Geometric}",
      journal = {arXiv e-prints},
         year = 2019,
        month = mar,
          eid = {arXiv:1903.02428},
        pages = {arXiv:1903.02428},
archivePrefix = {arXiv},
       eprint = {1903.02428},
 primaryClass = {cs.LG},
       adsurl = {https://ui.adsabs.harvard.edu/abs/2019arXiv190302428F}
}

@inproceedings{Okuta2017CuPyA,
  title={CuPy : A NumPy-Compatible Library for NVIDIA GPU Calculations},
  author={Ryosuke Okuta and Yuya Unno and Daisuke Nishino and Shohei Hido and Crissman},
  year={2017}
}

@inproceedings{he2016deep,
  title={Deep residual learning for image recognition},
  author={He, Kaiming and Zhang, Xiangyu and Ren, Shaoqing and Sun, Jian},
  booktitle={Proceedings of the IEEE conference on computer vision and pattern recognition},
  pages={770--778},
  year={2016}
}

@article{kalman-filter,
    author = "Fruhwirth, R.",
    title = "{Application of Kalman filtering to track and vertex fitting}",
    reportNumber = "HEPHY-PUB-87-503",
    doi = "10.1016/0168-9002(87)90887-4",
    journal = "Nucl. Instrum. Meth. A",
    volume = "262",
    pages = "444--450",
    year = "1987"
}

@inproceedings{10.3115/1072064.1072067,
author = {Chinchor, Nancy},
title = {MUC-4 Evaluation Metrics},
year = {1992},
isbn = {1558602739},
publisher = {Association for Computational Linguistics},
address = {USA},
url = {https://doi.org/10.3115/1072064.1072067},
doi = {10.3115/1072064.1072067},
booktitle = {Proceedings of the 4th Conference on Message Understanding},
pages = {22–29},
numpages = {8},
location = {McLean, Virginia},
series = {MUC4 ’92}
}

@misc{exatrkx,
title = {
HEP advanced tracking algorithms at the exascale},
year = {2019},
url={https://exatrkx.github.io/},
author = {The Exa.TrkX Collaboration},
}

@misc{heptrkx,
title = {
HEP advanced tracking algorithms with cross-cutting applications},
year = {2016},
url={https://heptrkx.github.io/},
author = {The HEP.TrkX Collaboration},
}

\end{document}